\documentclass[float,aps,prc,superscriptaddress,showpacs,twocolumn]{revtex4-1}
\usepackage{float}
\usepackage{color}
\usepackage{bm}
\usepackage{amsmath}
\usepackage{amssymb}
\usepackage{epsfig}
\usepackage[colorlinks,breaklinks]{hyperref}
\usepackage[titletoc]{appendix}
\usepackage{bbm}
\usepackage{multirow}
\usepackage{graphicx}

\begin{document}

\title{Machine learning approach to pattern recognition in nuclear dynamics\\ 
from the \textit{ab initio} symmetry-adapted no-core shell model}

\author{O.~M. Molchanov} \affiliation{Department of Physics and Astronomy, Louisiana State University, Baton Rouge, LA 70803, USA}

\author{K.~D. Launey} \affiliation{Department of Physics and Astronomy, Louisiana State University, Baton Rouge, LA 70803, USA}

\author{A. Mercenne} \affiliation{Department of Physics and Astronomy, Louisiana State University, Baton Rouge, LA 70803, USA}\affiliation{Center for Theoretical Physics, Sloane Physics Laboratory, Yale University, New Haven, CT 06520, USA}

\author{G.~H. Sargsyan}
\affiliation{Department of Physics and Astronomy, Louisiana State University, Baton Rouge, LA 70803, USA}

\author{T. Dytrych} \affiliation{Department of Physics and Astronomy, Louisiana State University, Baton Rouge, LA 70803, USA}\affiliation{Nuclear Physics Institute, Academy of Sciences of the Czech Republic, 25068 Rez, Czech Republic}

\author{J.~P. Draayer} \affiliation{Department of Physics and Astronomy, Louisiana State University, Baton Rouge, LA 70803, USA}


\date{\today}

\begin{abstract}
A novel machine learning approach is used to provide further insight into atomic nuclei and to detect orderly patterns amidst a vast data of large-scale calculations. 
The method utilizes a neural network that is trained on \textit{ab initio} results from the symmetry-adapted no-core shell model (SA-NCSM) for light nuclei. We show that the SA-NCSM, which expands \textit{ab initio} applications up to medium-mass nuclei by using dominant symmetries of nuclear dynamics, can reach heavier nuclei when coupled with the machine learning approach. In particular, we find that a neural network trained on  probability amplitudes for $s$-and $p$-shell nuclear wave functions not only predicts dominant configurations for heavier nuclei but in addition, when tested for the $^{20}$Ne ground state, it accurately reproduces the probability distribution. The nonnegligible configurations predicted by the network provide an important input to the SA-NCSM for reducing ultra-large model spaces to manageable sizes that can be, in turn, utilized in SA-NCSM calculations to obtain accurate observables. The neural network is capable of describing nuclear deformation and is used to track the shape evolution along the $^{20-42}$Mg isotopic chain, suggesting a  shape-coexistence that is more pronounced toward the very neutron-rich isotopes. We provide first descriptions of the structure and deformation of $^{24}$Si and $^{40}$Mg of interest to x-ray burst nucleosynthesis, and even of the extremely heavy nuclei such as $^{166,168}$Er and $^{236}$U, that build upon first principles considerations.
\end{abstract}

\maketitle


\section{Introduction}
Modeling atomic nuclei from first principles (or \textit{ab initio}) is a computationally demanding task. \textit{Ab initio} approaches use controlled approximations and interactions informed by the few-nucleon physics only, and therefore, they are  suitable to
determine reaction rates used in simulations of astrophysical processes, which involve short-lived nuclei that are impossible or difficult to be measured. With the symmetry-adapted no-core shell model (SA-NCSM) \cite{DytrychLDRWRBB20,LauneyDD16}, \textit{ab initio} calculations have been made possible up to the calcium region by utilizing symmetries that are inherent to atomic nuclei \cite{LauneyMD_ARNPS21}. However, reaching even heavier nuclei remains a challenge. The reason is that the number of basis states (model space size) in no-core shell models exponentially increases with the number of particles and the space they occupy. While the SA-NCSM drastically reduces the model space size based on an established symmetry-based prescription, this selection is validated through multiple simulations that ensure convergence of results. Therefore, novel computational approaches are needed to study challenging nuclear systems, such as $^{40}$Mg that has been suggested to have an effect on x-ray burst nucleosynthesis modeling \cite{Schatz13}, as well as nuclei in the lanthanide and actinide regions of interest to r-process simulations \cite{MumpowerSMA16}.   

In this paper, we use a machine learning (ML) approach based on a neural network with the goal to predict nonnegligible configurations to be used in SA-NCSM model spaces in a much less computationally demanding way. Indeed, a multilayer feedforward neural network can be used to approximate complicated functions by adding hidden units until it reaches a desired accuracy \cite{hornik_multilayer_1989}. 
Deep learning algorithms have emerged as a promising alternative to physics approaches  \cite{KarniadakisKL21} and have already been applied to nuclear physics, including
extrapolations of nuclear observables \cite{Negoita_2019,neufcourt,PhysRevC.100.054326}, machine-learning-based inversion of nuclear responses \cite{raghavan}, studies of the unitary limit \cite{kaspschak}, as well as optimization of experimental techniques and data analysis \cite{bedaque}.
Here, we use data from \textit{ab initio} SA-NCSM wave functions to train a network on selected light nuclei, in order to find ubiquitous patterns in nuclear dynamics and use these patterns to make predictions for heavier nuclei.

Earlier SA-NCSM results have shown the emergence of highly ordered patterns from first principles within different nuclei that relate to the dominance of only a few nuclear shapes in low-lying states that vibrate and rotate \cite{DytrychLDRWRBB20}. This suggests that a neural network can indeed be beneficial and by detecting these patterns it can inform us about nuclei that have not yet been modeled. The universality of these patterns ensures that the network can be applied across the nuclear chart, while retaining the properties of the training data. With a moderate size of training data, the network can identify the collection of negligible configurations that are eliminated from the SA-NCSM model space. This results in significantly fewer calculations needed to achieve convergence of results, thereby increasing the applicability of the model. Furthermore, for data training sets that are sufficiently large, we show that the network provides practically accurate predictions of dominant configurations in light and intermediate-mass nuclei, with probability amplitudes being in a close agreement with those calculated in the SA-NCSM, as evident in calculations of $^{4}$He and $^{20}$Ne. Remarkably, this is achieved in a tiny fraction of the time, without the need for full \textit{ab initio}  calculations.

Specifically, we explore the capabilities of the neural network and its efficacy in using information rooted in first principles to provide descriptions in nuclear regions where large-scale calculations are computationally demanding or impossible. We use the illustrative example of $^4$He to demonstrate that such a network is able to train on data from smaller model spaces to make accurate predictions for the same nucleus in larger model spaces. We compare the network predictions for the intermediate-mass nuclei, $^{20}$Ne, $^{24}$Si, and $^{28}$Mg, to available SA-NCSM calculations, and use the network to predict results in larger model spaces.
In addition, we study Mg isotopes between the proton and neutron driplines with a focus on the ground state evolution with neutron number, and provide insights into the shape coexistence phenomenon
Finally, in an application of the network to extreme cases of heavy nuclei, we provide predictions for dominant configurations in $^{166}$Er, $^{168}$Er, and $^{236}$U.

\section{Theoretical Framework}

\subsection{Symmetry-adapted no-core shell model}
The SA-NCSM provides \textit{ab initio} descriptions of nuclei \cite{LauneyDD16}. By taking advantage of inherent symmetries, it is able to eliminate negligible nuclear configurations based on their deformations and to achieve model spaces with a manageable size. This has allowed \textit{ab initio} descriptions for intermediate and medium-mass nuclei \cite{DytrychLDRWRBB20,LauneyDD16,LauneyMD_ARNPS21,Ruotsalainen19,PhysRevC.100.014322}. 
The SA-NCSM takes as input the interaction between the nucleons. In this paper, we utilize wave functions obtained with the NNLO$_{\rm opt}$ nucleon-nucleon (NN) potential \cite{Ekstrom13} derived in the chiral effective-field-theory framework.

The SA-NCSM many-body basis states build upon harmonic oscillator (HO) single-particle states and are labeled schematically as:
\begin{equation}
  |\mathfrak{a};N(\lambda\,  \mu)\kappa L; (S_p S_n )S; JM\rangle,
  \label{SAbasis}
\end{equation}
where $S_p$, $S_n$, and $S$ denote proton, neutron, and total intrinsic spins, respectively. $N$ is the total number of HO excitation quanta. The deformation is represented by $\lambda$ and  $\mu$ quantum numbers, which inform how prolate and oblate a state is, that is, $(\lambda\, 0)$ indicates a prolate deformation, $(0\, \mu)$ indicates an oblate deformation, whereas $(0\, 0)$ denotes a spherical shape \cite{LauneyDD16,LauneyMD_ARNPS21}. The label $\kappa$ distinguishes multiple occurrences of the same orbital momentum $L$ in a given ($\lambda$  $\mu$). The $L$ is coupled with $S$ to the total angular momentum $J$ and its projection $M$. The symbol $\mathfrak{a}$ schematically denotes the additional quantum numbers needed to specify the basis. 
 \begin{figure}[th]
\centering
\includegraphics[width=0.44\textwidth]{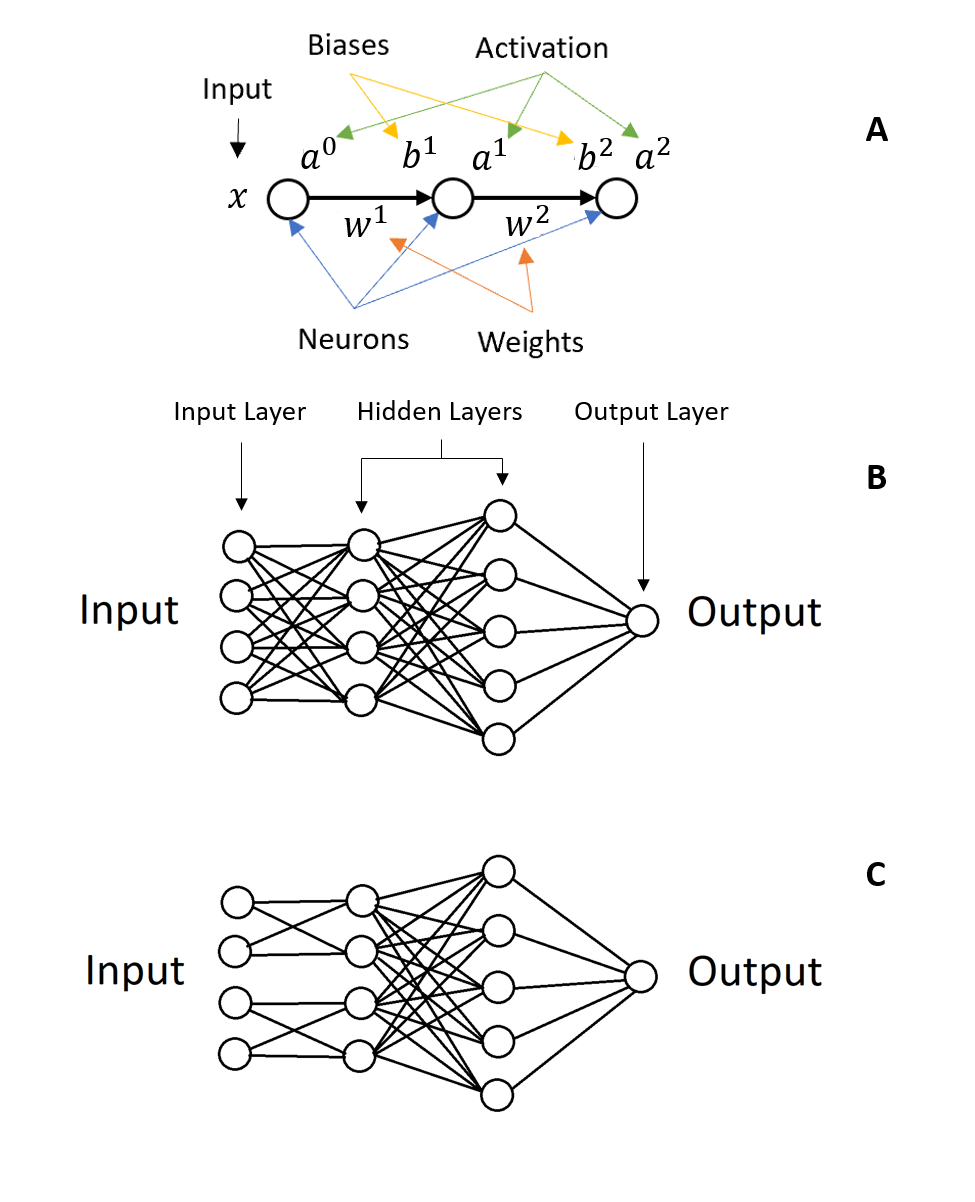}
\caption{(A) Feedforward neural network with interconnected neurons. Layers are labeled by index $i$; neurons are connected by weights $w^i$; each neuron has a bias $b^i$ and is assigned an activation function $a^i$. (B) A basic feedforward neural network with an input layer, two hidden layers, and an output layer with one neuron. In this network, every neuron in a layer is connected to every neuron in the previous layer. (C) A slightly more complicated network with segmented layers. Here, there are two groupings where two input neurons are exclusively connected to two hidden layer neurons but not the others.
}
\label{fig:network}
\end{figure}

Furthermore, the model space of the SA-NCSM is kept finite in size by introducing a maximum total number of HO excitation quanta $N_{\max}$ above the valence-shell configuration, that is, $N=0,\dots N_{\max}$, which defines the overall size of the model space. It also has an energy resolution of $\hbar \Omega$, defined by the energy of a single HO excitation. Both $N_{\max}$ and $\hbar \Omega$ are parameters of the basis, with converged results coinciding with those in the infinite model-space size ($N_{\max} \rightarrow \infty$) that are independent of $\hbar \Omega$. Calculations in a model space with a limited size (low $N_{\max}$) may result in enhanced deformations not fully developed. In this case, important basis states are not taken into account. Large enough $N_{\max}$ includes all necessary basis states at the cost of computational resources. In the SA-NCSM, the basis states are additionally selected above a given $N$ cutoff (labeled as $\langle N \rangle N_{\max}$). In this way, model subspaces above $N$ in the SA-NCSM include only non negligible configurations, and those are necessary to develop collective and clustering correlations.

\subsection{Network Structure}

A basic feedforward neural network, as shown in Fig.\ref{fig:network}a \& b, takes in inputs $\mathbf{x} \equiv \mathbf{a}^0$ in the input layer, and produces output $\mathbf{a}$ in the output layer. In each hidden layer $i$, the activation function of a neuron $k$ is determined from the activation functions of the previous layer's neurons $j$, $a^i_ k=f(\sum_{j}w^{i}_{jk}a^{i-1}_j+b^i_k)$, where the function $f$ is discussed below, and the network parameters $\mathbf{W}$ and $\mathbf{b}$ provide, respectively, the weights between two neurons and the bias associated with each neuron \cite{ketkar_feed_2017}. The value produced by the activation function is transmitted to the neurons in the next layer where this process happens again, until the output is reached. 
In this work, all layers except the output layer use the rectified linear unit (RLU) activation function for $f$. This RLU function returns the inputted value if it is positive, and it returns zero otherwise \cite{ketkar_feed_2017}. The output layer has a sigmoid activation function $S(x)$, where: 
\begin{equation}
  S(x) = \frac{1}{1 + e^{-x}}.  
\end{equation}

\begin{figure}[th]
(a) Poisson loss function\\
\includegraphics[width=\columnwidth]{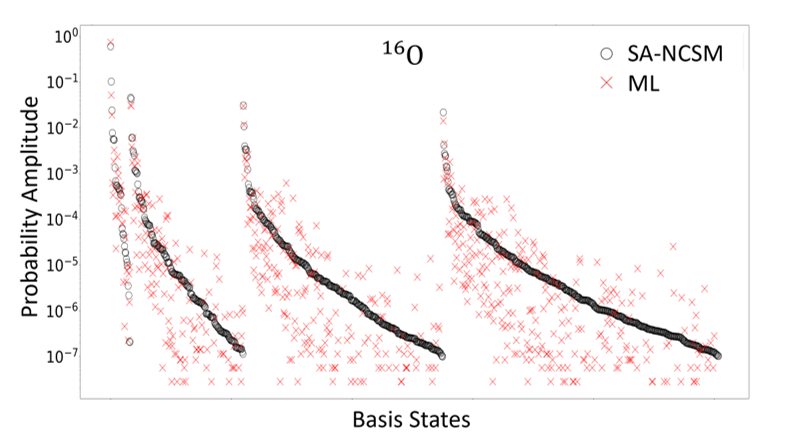}\\
\vspace{4pt}(b) Mean squared error loss function\\
\includegraphics[width=\columnwidth]{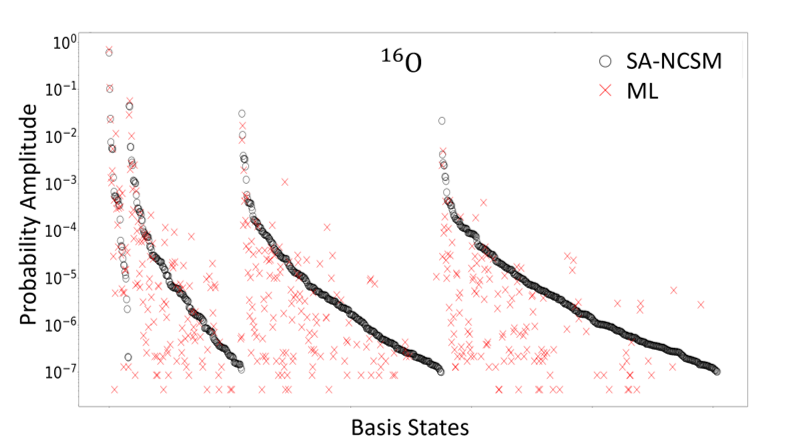}
\caption{
Probability amplitudes across basis configurations for $^{16}$O in $N_{\rm max}=8$ with $\hbar \Omega= 20$ MeV calculated in the large-scale many-body model (labeled as ``SA-NCSM") and predicted by the network (labeled as ``ML") trained on data for $^{4,8}$He, $^6$Li, $^8$Be, $^{10,12,14}$C, and $^{16}$O. Only basis states that have SA-NCSM probability amplitudes above $10^{-7}$ are shown. For states where the network predicts zero, no point is plotted for ``ML". (a) Predictions using Poisson loss function. (b) Predictions using the mean squared error as the loss function.
}
\label{fig:16O}
\end{figure}

To train a network, a set of data with inputs and their corresponding outputs is used to determine network parameters $\mathbf{W}$ and $\mathbf{b}$, for each connection and neuron, that minimize the difference between the results of the network (activation function of the output layer, $a_{\rm predicted}$), and the expected results (data outputs, $a_{\rm true}$). 
This difference defines a loss, which is thereby calculated by a loss function. In this work, we use the Poisson loss function,
\begin{equation}
 L = a_{\rm predicted} - a_{\rm true}  \ln(a_{\rm predicted}),
\end{equation}
which would prevent strict fitting to the true values in favor for an expected probabilistic pattern. For our case, using this instead of a standard mean squared error loss function resulted in data points that are more closely clustered around the expected values
and yielded a better fit when reproducing the training data (Fig. \ref{fig:16O}). 
The loss function $L$ is minimized using an optimizer \cite{ketkar_stochastic_2017}, which defines how the weights and biases are updated during training. The present approach utilizes Adam optimization, which is a stochastic gradient descent method with adaptive learning rates between parameters and carried momentum between updates \cite{kingma2017adam}.
In this study, the activation functions, loss function, and optimizer are chosen during test runs to ensure that the network fits the data well and have a capability of avoiding local minima. The training is run over 200 epochs and each epoch covers the entirety of the training set. In an effort to further avoid local minima, multiple networks may be individually trained to search for a network with nonzero and non-constant predictions for extreme cases, such as the heavy Er and U isotopes. 
The computational implementation of the network is achieved using the Keras deep learning application programming interfaces \cite{chollet2015keras}.

In this work, the neural network takes in a specific many-body basis configuration of a nuclear state and returns a value for the probability amplitude of that configuration. Since this network trains on data from SA-NCSM, it must have inputs relating to the parameters of the SA-NCSM model space. The presented results only use data with the NNLO$_{\rm opt}$ interaction to maintain consistency.

There are 11 input neurons. Each neuron has a single value as input. The inputs for the neurons are the number of protons $Z$, number of neutrons $A-Z$, total angular momentum $J$, excitations $N$, spin of protons $S_p$, spin of neutrons $S_n$, total spin $S$, $\lambda$, and $\mu$ according to Eq. (\ref{SAbasis}), as well $N_{\rm max}$ and  $\hbar \Omega$ (in MeV) basis parameters. However, for the purposes of reducing the scope of the data used by the network for this study, all of the cases have zero total angular momentum only. The inputs are entered as an array: 
\begin{equation}
  \left[Z,A-Z,N_{\rm max},\hbar \Omega,J,S_p,S_n,S,\tilde \lambda,\tilde \mu,N \right],
\end{equation}
where $\tilde \lambda=\lambda-\lambda_0$ and $\tilde \mu=\mu-\mu_0$ with $(\lambda_0\, \mu_0)$ being the configuration with the largest deformation and lowest intrinsic spin in the $N=0$ model subspace, also known as the leading SU(3) configuration, which is unique  for a given nucleus. This is based on a pattern we notice in the existing SA-NCSM results across nuclei, namely, the most dominant configurations often lie in the set 
\begin{eqnarray}
   N(\lambda\, \mu)=\{ && 0(\lambda_0\, \mu_0),\,
   \nonumber\\ 
   &&2(\lambda_0+2\,\mu_0),\, 2(\lambda_0-2\,  \mu_0+2),\, \nonumber\\
   &&4(\lambda_0+4\, \mu_0),\, 4(\lambda_0\, \mu_0+2), \, \nonumber\\
   &&6(\lambda_0+6\, \mu_0),\, 6(\lambda_0+2\, \mu_0+2),\dots \}  
   \label{pattern}
\end{eqnarray}
(see also Refs. \cite{LauneyDD16,DytrychLDRWRBB20}). By subtracting $(\lambda_0\, \mu_0)$, we expect the same pattern $\{ 0(0\,0),\, 2(2\, 0),\, 2(-2\, 2),\, 4(4\, 0),\, 4(0\,  2), \, 6(6\, 0), \, 6(2\, 2),\dots \} $ to be detected in all nuclei. For simplicity, we will drop the tilde notations from $(\lambda\,\mu)$ in further discussions of the neural network input.

\begin{figure}[th]
\includegraphics[width=0.9\columnwidth,trim={0.82cm 2.25cm 5.25cm 3.25cm},clip]{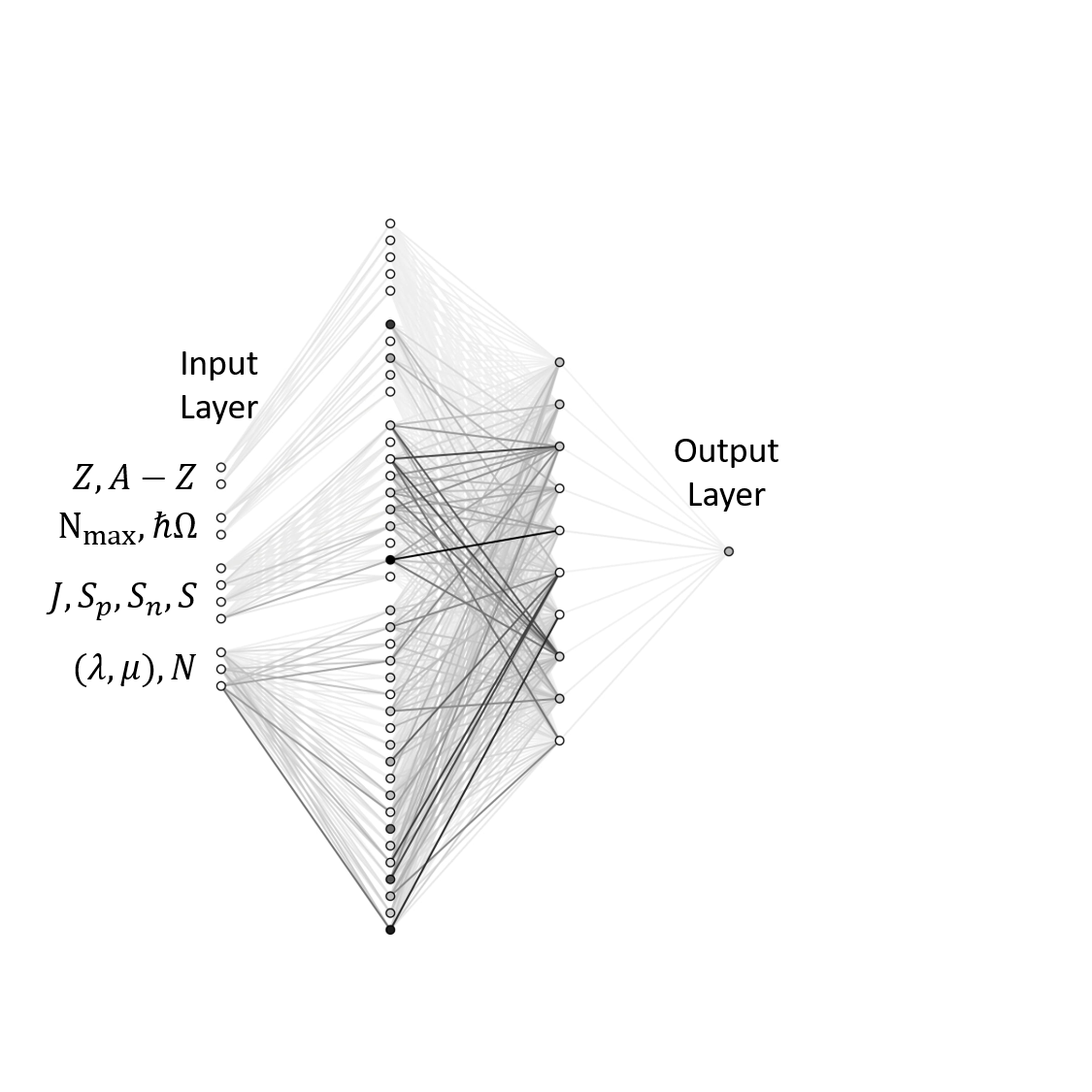}
\caption{
Neural network structure. Circles represent neurons and lines represent weights. Darker circles signify a larger absolute value of the bias. Darker lines signify a larger absolute value of the weight. Input neurons are in the left-most layer. There are two hidden layers. The input layer and first hidden layer have their connections segmented into groups. The input labels are shown next to the input neurons in downward order. The second hidden layer and output layer connect to all neurons before them. The output is the right-most neuron.
}
\label{fig:NN}
\end{figure}

As mentioned above, in a basic neural network, when one layer serves as input to a second layer, it is common for each neuron in the second layer to be connected to every single neuron in the first layer by some set of weights (Fig. \ref{fig:network}b). For applications to SA-NCSM, we find that a network with a segmented first hidden layer, as schematically shown in Fig. \ref{fig:network}c, provides important improvements. Since the input layer includes different kinds of information about the nucleus, including particle numbers ($Z$, $A-Z$), basis parameters ($N_{\rm max}$ and $\hbar \Omega$), spins of a nuclear state ($J$, $S_p$, $S_n$, and $S$), and its deformation, ($N$, $\lambda$, and $\mu$), it is beneficial to group these inputs into different sets (Fig. \ref{fig:NN}). Each of these sets will then have their own segment of the first hidden layer that they connect to. This segment will not use any values from the other sets as input. For each of these sets we know that information within has strong relationships that affects the output. Segmenting the first layer, therefore, forces the network to detect correlations between these inputs independently from other inputs.

The first hidden layer has some segments with more neurons to reflect the  increased complexity expected for the relationship between the output and the spins, and more so, the deformation $N(\lambda \,\mu)$. 
The second hidden layer has 10 neurons, but connects to all of the first hidden layer. The number of neurons was chosen in an effort to have the number of parameters be about 0.01 of the number of data points in the training set (Table \ref{tab:trainingset}). Finally, there is one output neuron giving the probability of the configuration. The network is used to give a prediction for the probability amplitudes for all possible configurations for a nucleus with given $J$ and for given $\hbar \Omega$ and $N_{\rm max}$, which are normalized afterwards. The overall structure of the network can be visualized with Fig. \ref{fig:NN}. 
\begin{table}[H]
\centering
\caption{ Training and validation data sets. A data set for given $N_{\rm max}$ and $\hbar \Omega$ includes tens through hundreds to thousands of configurations. Complete model spaces are denoted by $N_{\rm max}$, whereas selected model spaces are denoted as $\langle N \rangle N_{\rm max}$ (e.g., for $^{20}$Ne, $\langle 2 \rangle 10$ indicates the use of all basis states up to $N=2$ and selected basis states for $N=4$, 6, 8, and 10). Also shown is the dimension (number of all basis states) of the largest $J=0$ model space listed for each nucleus.}
\begin{tabular}{|c|c|c|c|}
\hline
Nucleus & $N_{\rm max}$ & $\hbar \Omega$ (MeV) & Dimension \\
\hline
\multicolumn{4}{|c|}{Training data}\\\hline
$^4$He & 6,8,10,12,14 & 22 & 5.80$\times 10^4$\\
$^8$He & 6 & 15 & 3.18$\times 10^4$ \\
$^6$Li & 8 & 15 & 6.78$\times 10^4$\\
$^8$Be & 8,10,12,$\langle 8 \rangle 14$ & 15 & 1.53$\times 10^8$\\
$^{10}$C & 8 & 15 & 4.44$\times 10^6$ \\
$^{12}$C & 6,$\langle 6 \rangle 10$,$\langle 6 \rangle 12$ & 15 & 1.90$\times 10^9$ \\
$^{14}$C & 6,$\langle 6 \rangle 10$ & 15 & 4.03$\times 10^8$ \\
$^{16}$O & 8 & 20,25 & 3.01$\times 10^7$\\ \hline
\multicolumn{4}{|c|}{Validation data}\\\hline
$^{20}$Ne & $\langle 4 \rangle 6$,$\langle 4 \rangle 8$,$\langle 2 \rangle 10$  & 15 & 7.18$\times 10^{10}$\\
$^{28}$Mg & $\langle 0 \rangle 6$ & 15 & 9.77$\times 10^9$\\
$^{24}$Si & 4 & 20 & 2.59$\times 10^7$ \\
\hline
\end{tabular}
  \label{tab:trainingset}
\end{table}

\section{Results and Discussion}
Results discussed in this paper, except those for $^4$He, are presented for a training set that  includes only the $s$- and $p$-shell nuclei. This includes the $s$-shell nucleus $^4$He  and $p$-shell nuclei $^6$Li, $^8$He, 
$^8$Be, $^{10,12,14}$C, and $^{16}$O, as listed in Table \ref{tab:trainingset}. The neural network predicts probability amplitudes for a ground state wave function, which are subsequently normalized to one. In this section, we present network validation (for $^4$He and $^{20}$Ne), as well as network predictions in heavier nuclei ($^{24}$Si, $^{28}$Mg, $^{166,168}$Er and $^{236}$U), together with a shape evolution  along the Mg isotopic chain.

\subsection{Network validation}

\begin{figure}[th]
\centering
\includegraphics[width=1.07\columnwidth]{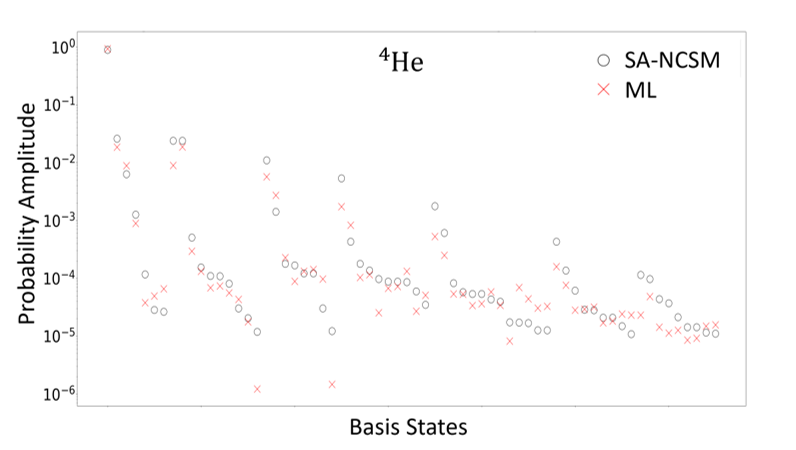}
\caption{Probability amplitudes across basis configurations for $^4$He in $N_{\rm max}=16$ calculated in the large-scale many-body model (labeled as ``SA-NCSM") and predicted by the network (labeled as ``ML") trained on data for $^4$He in $N_{\rm max}=8$, 10, 12, and 14. Only basis states with SA-NCSM probability amplitudes above $10^{-5}$ are shown. For states where the network predicted zero, no point is plotted for ``ML".
}
\label{fig:4He}
\end{figure}

We study  
the capability of the neural network to predict larger model spaces by training on smaller ones, and show an example for the $^4$He ground state (Fig. \ref{fig:4He}). In this case, we expect that $^4$He develops its most significant structures in lower $N_{\rm max}$ spaces and hence, no new patterns are expected as one goes from $N_{\rm max}=12$ with 22,716 basis states to $N_{\rm max}=14$ with 58,080 basis states. Indeed, the results show that the network has the capability to make close predictions of probabilities of basis states for data similar to the training data.

\begin{figure}[t]
\centering
\includegraphics[width=0.49\textwidth]{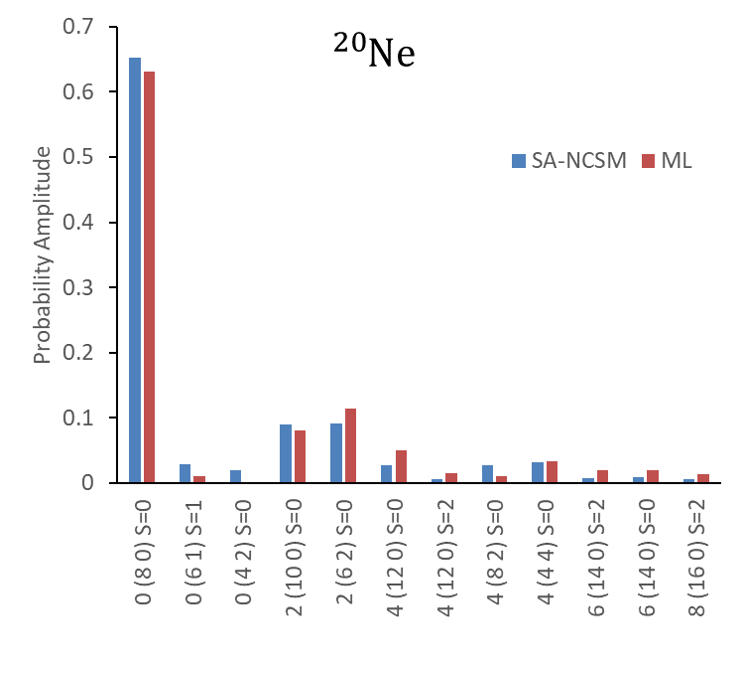}
\caption{Probability amplitudes of dominant configurations predicted for the $^{20}$Ne ground state in $N_{\rm max}=10$ by the network trained on $^{4,8}$He, $^6$Li, $^8$Be, $^{10,12,14}$C, and $^{16}$O, as compared to the SA-NCSM calculations. The configurations are labeled by $N(\lambda\, \mu)$ and shown for the largest SA-NCSM probability amplitudes $\ge 1\%$.The configurations are labeled by $N(\lambda\, \mu)S$ and, for each $N$, listed left to right with decreasing deformation. 
}
\label{fig:20Ne}
\end{figure}

The most important objective of using machine learning in the present approach is to predict information about heavier nuclei using \textit{ab initio} calculations for light nuclei. As an illustrative example we validate the network with results for the intermediate-mass nucleus $^{20}$Ne from calculations of $s$- and $p$-shell light nuclei for $N_{\rm max}=6,8,$ and 10 (Table \ref{tab:trainingset}). We show predictions for the largest model space $N_{\rm max}=10$ (Fig. \ref{fig:20Ne}). What we predict is
an $sd$-shell nucleus, $^{20}$Ne, which is not included in the training set. We 
utilize a $N_{\rm max}=10$ model space that consists of about $10^{12}$ basis states that is currently 
infeasible, but can be reduced  
to several million SA basis states
with a solution in the SA-NCSM \cite{DytrychLDRWRBB20}. 
Remarkably, we find that the network predictions for the most significant configurations
in the $^{20}$Ne ground state follow 
the same pattern as the one revealed by the SA-NCSM and is associated with an important and widely spread feature of nuclear dynamics, namely, vibrations of equilibrium shapes \cite{DytrychLDRWRBB20}. 
Note that for $^{20}$Ne, $(\lambda_0 \, \mu_0)$ is $(8\, 0)$, and the dominant configurations of Eq. (\ref{pattern}), $0(8\, 0)$, $2(10\, 0)$, $2(6\, 2)$, $4(12\, 0)$,  $\dots$ are clearly revealed, as shown in Fig. \ref{fig:20Ne}. Even beyond, the network results are in a very close agreement with the SA-NCSM probability amplitudes. This implies that once the network is trained on light nuclei, it can efficiently provide probability amplitudes for a deformation-based wave function and related information, e.g., shape dominance and coexistence along with moments of inertia, without the need for large-scale computationally intensive calculations. It can also provide an upper estimate for E2 transition strengths. This result is important as it shows that it is not necessary to include $sd$-shell nuclei in the training set to get good predictions in the $sd$-shell region. However, we find that it is imperative to use sufficiently large training data sets, as tests with networks trained only on a few nuclei result in larger deviations, but increasing the volume of training data remedies this.
\begin{figure}[th]
\centering
(a) {\small $N_{\rm max}=4$}\\
\includegraphics[width=1.05\columnwidth,height=2.3in]{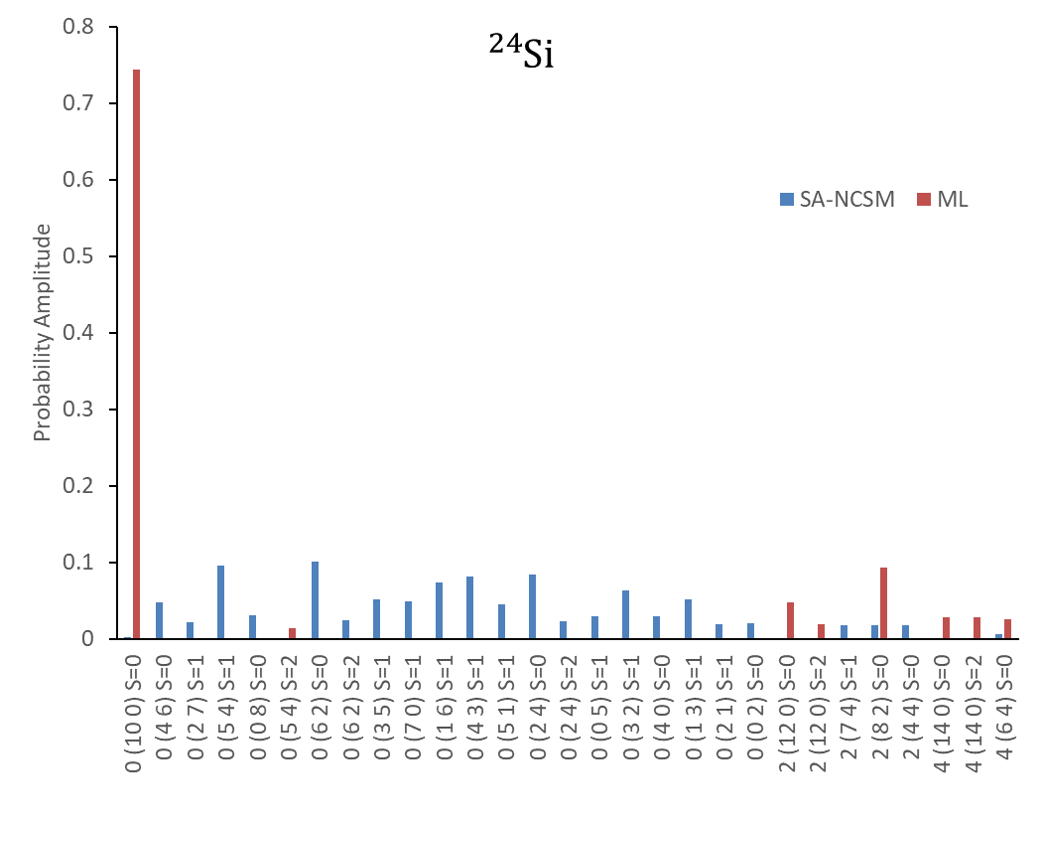}\\
\vspace{4pt}(b) {\small $N_{\rm max}=12$}\\
\includegraphics[width=\columnwidth,height=2.2in]{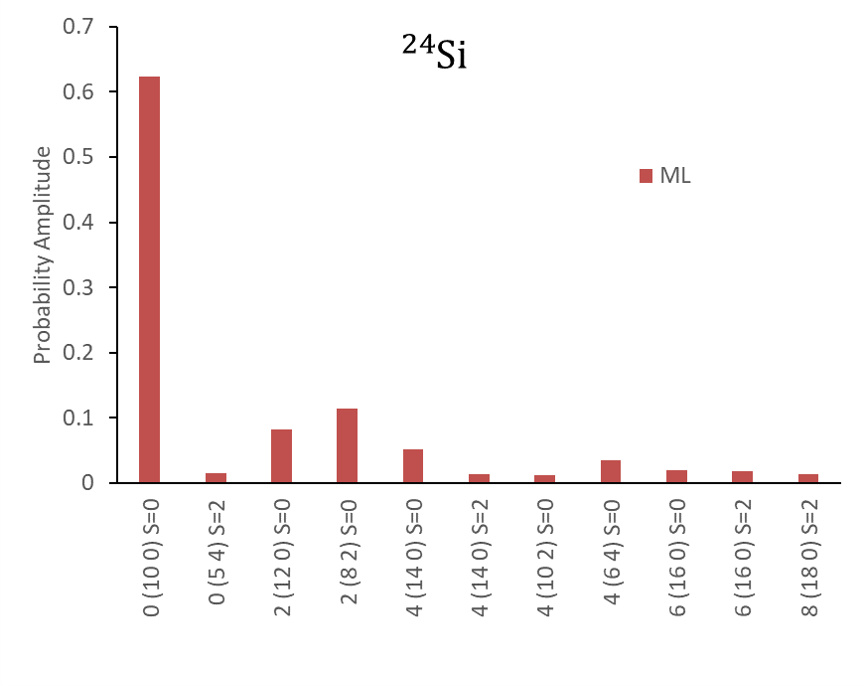}
\caption{
Probability amplitudes of dominant configurations 
predicted for the $^{24}$Si ground state by the network (``ML") trained on $^{4,8}$He, $^6$Li, $^8$Be, $^{10,12,14}$C, and $^{16}$O, and compared to available SA-NCSM calculations. (a) $N_{\rm max}=4$ model space; (b) $N_{\rm max}=12$ model space, with configurations with probability amplitudes $\ge 1\%$.  The configurations are labeled by $N(\lambda\, \mu)S$ and, for each $N$, listed left to right with decreasing deformation. 
}
\label{fig:24Si}
\end{figure}

It is interesting to examine the weights and biases of the network itself (Fig. \ref{fig:NN}). From the proton and neutron inputs, the outgoing weights are relatively small. This is expected since the dominant configuration $(\lambda_0\, \mu_0)$ is subtracted from the $(\lambda\,  \mu)$ before it enters as an input, as discussed above. This dominant configuration depends heavily on the numbers of these particles. If this subtraction was not implemented, then we would expect more heavy weights from those neurons. In our case, the network is building on top of the dominant ($\lambda_0$  $\mu_0$), so while the proton and neutron numbers still affect the pattern, the network uses them for fine tuning since the pattern is indeed similar across different nuclei. The largest deformation generally comes from a zero spin state, so this is likely reflected in how the total spin $S$ neuron has the heaviest weight within the spin segment of the network. The pattern previously noted in Eq. (\ref{pattern}) also depends on $N$, thus the network has multiple relatively heavier weights from the $N$ input neuron. The heaviest weights in the network are between the two hidden layers, most notably from the spins and the deformation segments in the first hidden layer, showing that the network reflects the complexity associated with these quantum numbers. The biases mainly vary in the first hidden layer. Both the biases in the second hidden and the weights connecting to the output neuron have a smaller range. This suggests that the network calculates the final probability by taking into account a number of similar contributions rather than arising from a single main pattern stored in one neuron in the last hidden layer.

\subsection{Predictions for \textit{sd}-shell nuclei}

We present another example for an $sd$-shell nucleus, the ground state of $^{24}$Si, where we compare no-core shell-model results in small model spaces to the network prediction (Fig. \ref{fig:24Si}). The %
\begin{figure}[th]
(a) {\small $N_{\rm max}=6$}\\
\includegraphics[width=1.03\columnwidth,height=2in]{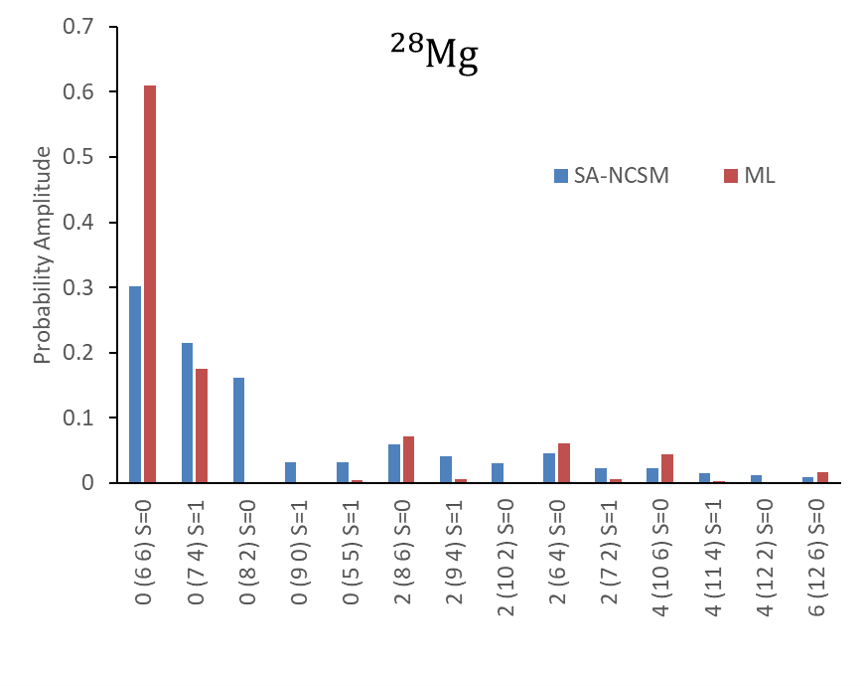}\\
\vspace{4pt}(b) {\small $N_{\rm max}=12$}\\
\includegraphics[width=1.03\columnwidth,height=2in]{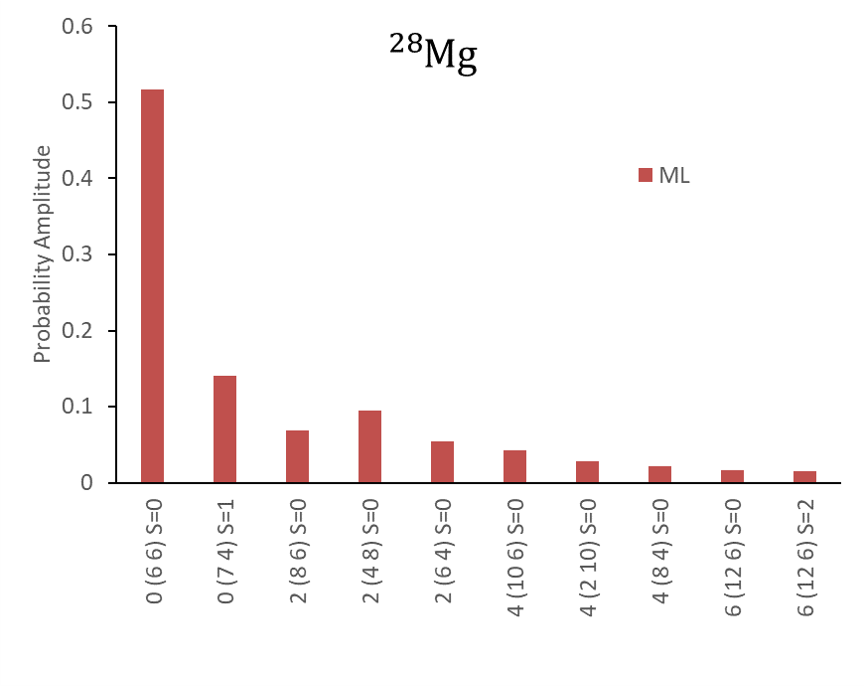}
\caption{
Probability amplitudes of dominant configurations predicted for the $^{28}$Mg ground state by the network (``ML") trained on $^{4,8}$He, $^6$Li, $^8$Be, $^{10,12,14}$C, and $^{16}$O, and compared to available SA-NCSM calculations. (a) $N_{\rm max}=6$ model space; (b) $N_{\rm max}=12$ model space, with configurations with probability amplitudes that are $\ge 1\%$.  The configurations are labeled by $N(\lambda\, \mu)S$ and, for each $N$, listed left to right with decreasing deformation.
}
\label{fig:28Mg}
\end{figure}
results largely differ when the SA-NCSM calculations are limited to $N_{\rm max}=4$. In this very small 
model space, the dominant deformations cannot fully develop. However, larger 
model spaces are very computationally intensive and prohibitive for many models (the complete space in 
$N_{\rm max}=6$ has dimension of $\sim 10^{11}$). The network predicts $(10\, 0)$ to be the most dominant configuration even within a 
limited model space, and in addition recognizes large configurations in $N=2$ and $N=4$ from the pattern of Eq. (\ref{pattern}) (Fig. \ref{fig:24Si}a). In contrast, the limited $N_{\rm max}=4$ SA-NCSM calculations, as expected, fail to account for such collective correlations. The reason is that,  within an $N_{\rm max}$ model space,  the no-core shell model minimizes the ground state energy (according to the variational principal) and to achieve this, introduces spurious (less deformed) configurations that favor lower energies at the given model space cutoff. The network, however, builds upon the collective correlations and clearly reveals some of them even in relatively small model spaces. In larger model spaces, the network suggests a slightly smaller contribution of the predominant shape as other shapes become slightly more important (Fig. \ref{fig:24Si}b).
The network results point to physics that is similar to neighboring nuclei and clearly omitted in no-core shell-model calculations in limited model spaces. This information is critical for the construction of a selected SA-NCSM model space for $^{24}$Si, which, as a next step, will be used as input to large-scale SA-NCSM calculations for first \textit{ab initio} predictions of various observables, such as energy spectrum, radii, and reaction rates, for low-lying states in $^{24}$Si that currently cannot be measured directly.    

For an example of predicting an even heavier nuclei, we study the challenging $^{28}$Mg (Fig. \ref{fig:28Mg}), which has been suggested to lie in the so-called island of inversion largely affected by the higher $pf$ shell \cite{PhysRevC.100.014322}. 
In this case, the network is also capable of detecting the significant configurations, and even reproduce the probability amplitudes to a good degree. In particular, for $N_{\rm max}=6$, the most dominant configuration is predicted by the network, but with larger probability as compared to the SA-NCSM calculations, while another important configuration, $0 (8\, 2) S=0$, is drastically underestimated (with probability amplitude of only 0.084\%). The second dominant shape $0(7\, 4) S=1$, when combined across proton and neutron spins, is also in a reasonable agreement. It is interesting that the network recognizes the most dominant $S=0$ mode, and especially the most dominant $S=1$ modes which are often suppressed within the light nuclei in the training set. The dominant configurations for $N=2$ and $N=4$ are also found to be well reproduced, but here again some slightly less significant modes are neglected entirely according to the network output. The network is then applied to the larger $N_{\rm max}=12$ model space (Fig. \ref{fig:28Mg}b), where the results reveal a very similar pattern as the one predicted for $N_{\rm max}=6$, with a slightly smaller $(6\,6) S=0$ contribution. Hence, the network suggests a predominance of a triaxial shape in $^{28}$Mg, the same one detected at the smaller model space, and some admixture with an oblate shape.
\begin{figure}[H]
\centering
\includegraphics[width=0.8\columnwidth]{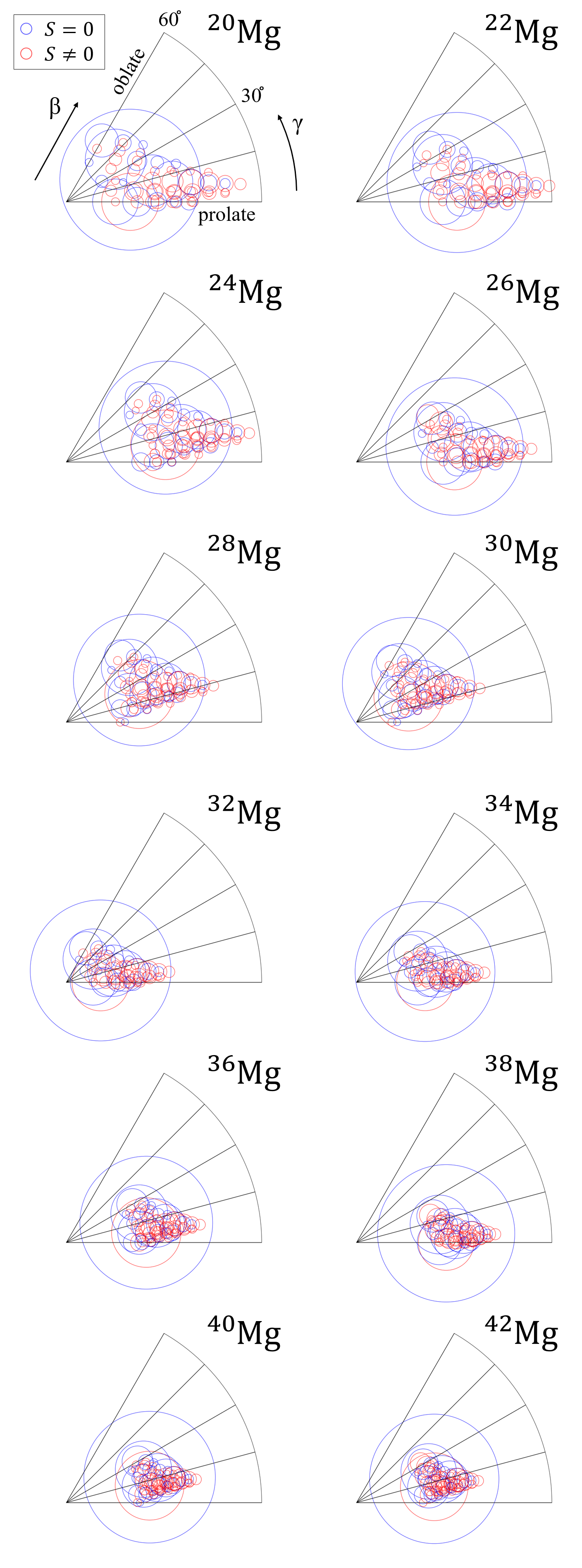}
\caption{Probability distribution (specified by the area of the circles) across the deformation $\beta$ and triaxiality $\gamma$ shape parameters for various isotopes of Mg, in $N_{\rm max}=12$, trained on $^{4,8}$He, $^6$Li, $^8$Be, $^{10,12,14}$C, and $^{16}$O. The total spin that contributes most to each configuration is also shown for $S=0$ (blue) and nonzero $S$ (red).}
\label{fig:isotopesMg}
\end{figure}
\begin{figure}[H]
\center(a)\\
\includegraphics[width=\columnwidth]{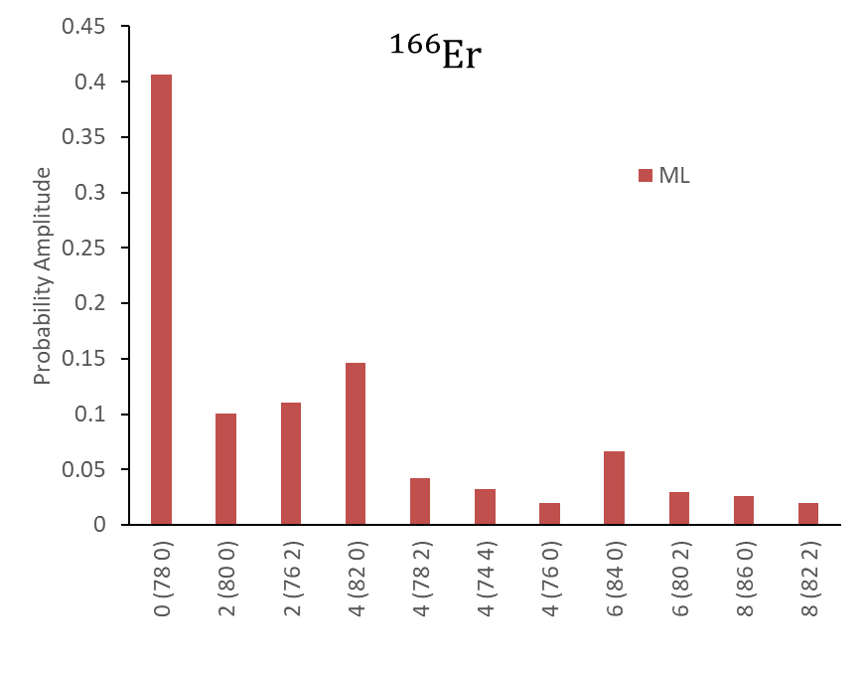}\\
\vspace{4pt}\center(b)\\
\includegraphics[width=\columnwidth]{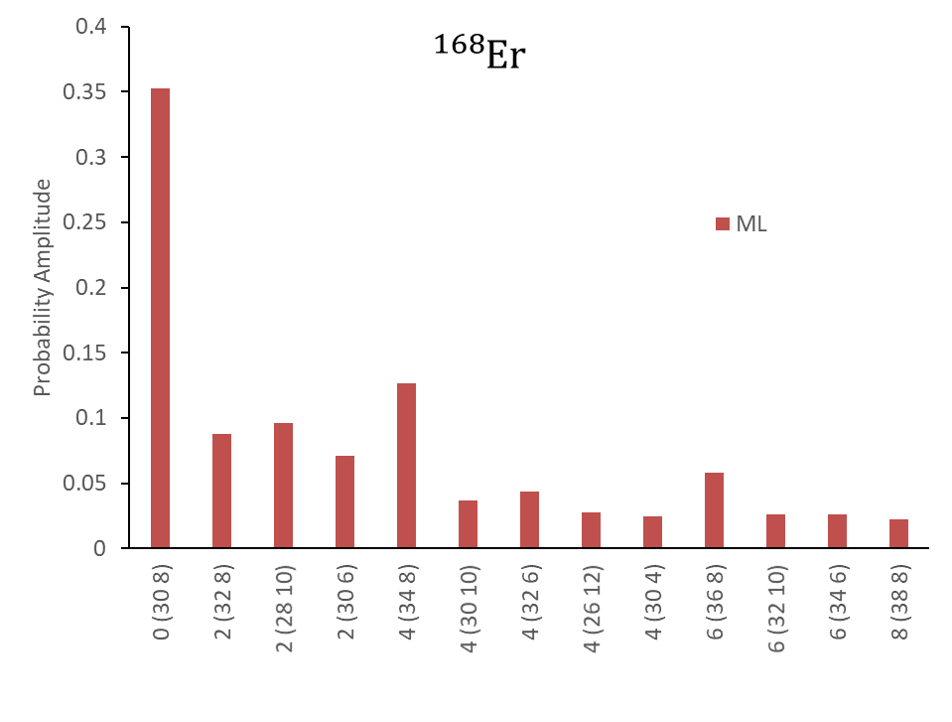}\\
\vspace{4pt}\center(c)\\
\includegraphics[width=\columnwidth]{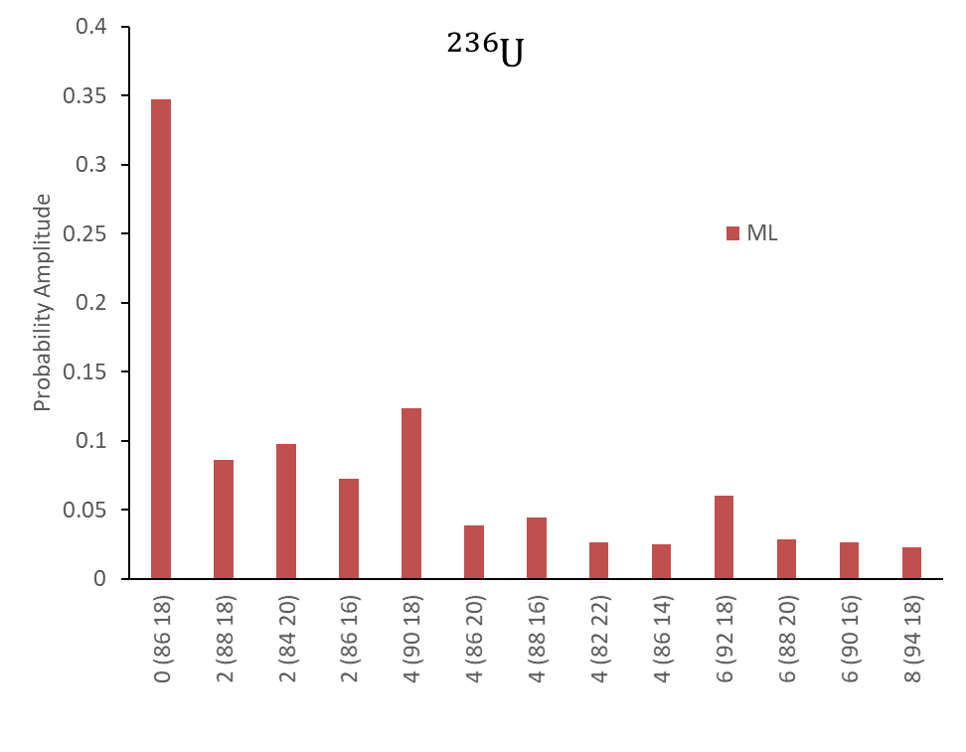}
\caption{Largest probability amplitudes ($\ge 1\%$) across basis configurations predicted by the network for (a) $^{166}$Er in $N_{\rm max}=12$, (b) $^{168}$Er in $N_{\rm max}=12$, and (c) $^{236}$U in $N_{\rm max}=12$. Network trained on $^{4,8}$He, $^6$Li, $^8$Be, $^{10,12,14}$C, and $^{16}$O.}
\label{fig:ErU}
\end{figure}

We expect that including more isotopes in the training set with different numbers of protons and neutrons would further improve the network accuracy for both $^{24}$Si and $^{28}$Mg, since a large part of the current training set consists of  nuclei with equal numbers of protons and neutrons.

\subsection{Shape evolution in Mg isotopes}

As an important outcome, we utilize the neural network to study shape evolution in Mg isotopes, from the proton-rich $^{20}$Mg to the $^{40,42}$Mg near the neutron drip line. Since these are open-shell nuclei in the intermediate- and medium-mass region, these systems pose a challenge to \textit{ab initio} theory. We consider even-mass nuclei only, since all of the training data is for even-mass nuclei. For each isotope, the $(\lambda \, \mu)$ quantum numbers of the basis 
states can be directly linked to the deformation $\beta$ and triaxiality $\gamma$ shape parameters \cite{BohrMottelson69,CastanosDL88} (Fig. \ref{fig:isotopesMg}). All of the isotopes favor prolate shapes except $^{28}$Mg, which is triaxial, and $^{30}$Mg, which is oblate. For example, the $^{26}$Mg deformation distribution  predicted by the network closely resembles the E2 transition density previously calculated in the constrained Hartree-Fock-Bogoliubov plus local quasiparticle random-phase approximation method  \cite{PhysRevC.83.014321}. In all cases, there is a predominant shape, which has the expected $(\lambda_0 \, \mu_0)$ deformation and intrinsic spin $S=0$, followed by another shape with $(\lambda_0+1 \, \mu_0-2)$ $S=1$. In general, a large ratio of the probability amplitudes for the secondary shape relative to the dominant shape indicates a strong interplay of the two shapes, pointing to a  shape coexistence in the low-lying energy spectrum. It is interesting that the network suggests multiple cases where the shape coexistence becomes pronounced, namely, the neutron-rich
$^{28}$Mg, $^{30}$Mg, $^{36}$Mg, $^{40}$Mg, and $^{42}$Mg. This may be a result of reduced spin-orbit interaction toward the drip line. In addition to these two shapes, there are other $S=0$ configurations that appear important, however, they follow the pattern of Eq. (\ref{pattern}) and are associated with vibrations of the dominant shape rather than the existence of a new equilibrium shape (cf. Ref. \cite{DytrychLDRWRBB20}).

\subsection{Neural network results for the actinide and lanthanide region}

Using the neural network, it is even possible to make predictions for extremely heavy nuclei, as shown in Fig. \ref{fig:ErU} for $^{166,168}$Er and $^{236}$U. For such nuclei, full \textit{ab initio} calculations are definitely not feasible. Yet, the network is able to suggest a very similar pattern in these nuclei as the one observed in light and medium-mass nuclei. There is only a small difference, namely, vibrations of the nuclear shape along the $x$ and $y$ axes are comparable to those in the $z$ direction at $N=2$. Nonetheless, the network clearly recognizes a set of dominant configurations that resembles the one suggested by earlier studies with schematic interactions \cite{BahriR00,Rowe13,TroltenierDHC94}. For $^{166}$Er, Ref. \cite{BahriR00} suggests the choice of $(\lambda_0\, \mu_0)=(78\,0) $ with all spins assumed to be zero, which has reproduced E2 experimental data. For $^{168}$Er and $^{236}$U, we use the leading $0(\lambda_0\, \mu_0)$ configurations as done for the lighter nuclei presented above, namely, $0(30\, 8)$ for $^{168}$Er and $0(86\, 18)$ for $^{236}$U.
We note that a different $(\lambda_0\, \mu_0)$ is suggested in Ref. \cite{TroltenierDHC94} based on the use of pseudo-SU(3). Comparing the networks predictions with the results of these earlier papers, we find that the network reproduces the dominant configurations and the overall pattern, while future studies of B(E2) strengths based on the neural network results will provide insight on the interplay of vibrations in the $z$, $x$, and $y$ directions in these nuclei. As mentioned above, the network is trained on nuclei that are very light compared the Er and U isotopes, so we expect that adding significantly more data from much heavier nuclei, e.g. SA-NCSM calculations around mass $A=50$, will improve the network predictions at this scale, since increasing the volume and diversity of training data is indeed beneficial.

\section{Conclusion}

In this study, we demonstrate that a neural network provides an efficient new approach to modeling nuclei, from light to heavy mass, by using first principles input. In particular, we showed that the network was able to train on data from a single nucleus to make accurate predictions of dominant configurations for larger $N_{\rm max}$ model spaces, as in the case of $^4$He. Training on multiple $s$- and $p$-shell nuclei allowed the network to accurately predict the dominant configurations for the $sd$-shell 
nucleus of $^{20}$Ne. This validates the network and suggests that the network is not  limited to predicting extremely similar nuclei.

In addition, the network results reasonably agreed with existing \textit{ab initio} SA-NCSM calculations for $^{28}$Mg, and we found that the network was especially suitable for identifying the nonnegligible configurations, which, in turn, could be used to inform the selection of model spaces for large-scale SA-NCSM calculations of various observables, including energies, radii, and electromagnetic moments and transitions.   
We further showed, for the illustrative example of $^{24}$Si, that the network was capable of detecting the important collective
correlations even in smaller model spaces where shell-model calculations fail to account for those.

As a notable outcome, the neural network was capable of detecting a ubiquitous feature of nuclear dynamics, namely, vibrations of equilibrium shapes, thereby applicable across the nuclear chart. This is important, since the network is not used to extrapolate to regions of different physics compared to the training data set.

The neural network was utilized to study the interesting phenomenon of shape coexistence and even to reach heavy nuclei. In particular,  the network results for different Mg isotopes were used to study the deformation distribution within each isotope, suggesting an interplay of two shapes as recognized in the neutron-rich
$^{28,30}$Mg, $^{36}$Mg, and $^{40,42}$Mg. Furthermore, the nature of the network allowed us to apply the approach to extreme cases that may never be accessible by \textit{ab initio} modeling, such as $^{166,168}$Er and $^{236}$U. Remarkably, this was accomplished with  training on $s$- and $p$-shell nuclei only. While adding $sd$-shell nuclei to the the training set will help the network achieve more accurate predictions, the lighter nuclei appear sufficient for the network to detect the patterns that can be used for these significantly heavier nuclei cases. We note that the network results remain to be tested against other observables, such as E2 transitions. 

We emphasize two important features of the network. We found that substantially increasing the volume of the training data set led to better predictive capability, which suggests that, as more SA-NCSM calculations are performed and added to the training set, the training data will become richer and hence, will increase the network predictive power. 
Another feature of the neural network is that the minimization procedure depends on a seed for a random number generator used to update base weights and biases. Hence, the loss function can reach a local minimum, preventing the network from finding the best fit to the patterns presented. Within a given run, there is an innate variability of how the weights and biases are chosen.
Hence, the results could be different even with two networks trained on the same data. To resolve this, we create multiple networks that are trained on the same data subset, and select the one with the lowest loss. This network is then used to continue training with the complete data set. Alternatively, multiple iterations may be performed, with the goal to reach the global minimum. Such a procedure will result in improved predictions and is part of ongoing work.

In short, we construct a novel machine learning approach that, coupled with large-scale \textit{ab initio} SA-NCSM calculations, provides further insight into atomic nuclei, and is capable of detecting orderly patterns amidst a vast data of large-scale calculations. This approach is ideal for studies and predictions of dominant shapes across the nuclear chart.

\begin{acknowledgments}
We acknowledge useful discussions with Manos Chatzopoulos. This work was supported in part by the U.S. National Science Foundation  (PHY-1913728), U.S. Department of Energy (DE-SC0019521), SURA, and the Czech Science Foundation (16-16772S).  It benefited from high performance computational resources provided by LSU (www.hpc.lsu.edu),  the National Energy Research Scientific Computing Center (NERSC), a U.S. Department of Energy Office of Science User Facility operated under Contract No. DE-AC02-05CH11231, as well as the Frontera computing project at the Texas Advanced Computing Center,  made possible by National Science Foundation award OAC-1818253.
\end{acknowledgments}

\bibliography{NucResPaper}

\end{document}